\documentclass[manuscript,screen]{acmart}
\AtBeginDocument{%
  }

\usepackage{xfrac}
\usepackage{flushend}
\usepackage{amsmath}
\usepackage{mathtools}
\usepackage{balance}
\usepackage{nicefrac}
\usepackage{multirow}
\usepackage{subcaption}

\usepackage{enumitem}
\usepackage{algorithm}
\usepackage[noend]{algpseudocode}

\setcopyright{acmcopyright}
\acmDOI{XXXXXXX.XXXXXXX}

\renewcommand\footnotetextcopyrightpermission[1]{}

\makeatletter
\newcommand{\algmargin}{\the\ALG@thistlm}
\makeatother
\newlength{\whilewidth}
\algdef{SE}[parWHILE]{parWhile}{EndparWhile}[1]
{\parbox[t]{\dimexpr\linewidth-\algmargin}{%
		\hangindent\whilewidth\strut\algorithmicwhile\ #1\ \algorithmicdo\strut}}{\algorithmicend\ \algorithmicwhile}%
\algdef{SE}[SUBALG]{Indent}{EndIndent}{}{\algorithmicend\ }%
\algtext*{Indent}
\algtext*{EndIndent}
\algnewcommand{\parState}[1]{\State%
	\parbox[t]{\dimexpr\linewidth-\algmargin}{\strut #1\strut}}
\makeatother
\DeclareMathAlphabet{\mathmybb}{U}{bbold}{m}{n}

\newtheorem*{lemma*}{Lemma}

\newcommand{\sref}[2]{\hyperref[#2]{#1 \ref{#2}}}

\begin{document}

\title[Dynamic Incentive Allocation for City-scale Deep Decarbonization]{Dynamic Incentive Allocation for City-scale Deep Decarbonization}

\author{Anupama Sitaraman}
\affiliation{%
  \institution{Carnegie Mellon University}
    \country{USA}
}

\author{Adam Lechowicz}
\affiliation{%
  \institution{University of Massachusetts Amherst}
    \country{USA}
}

\author{Noman Bashir}
\affiliation{%
  \institution{Massachusetts Institute of Technology}
    \country{USA}
}

\author{Xutong Liu}
\affiliation{%
  \institution{Carnegie Mellon University}
    \country{USA}
}

\author{Mohammad Hajiesmaili}
\affiliation{%
  \institution{University of Massachusetts Amherst}
    \country{USA}
}

\author{Prashant Shenoy}
\affiliation{%
  \institution{University of Massachusetts Amherst}
    \country{USA}
}

\renewcommand{\shortauthors}{Sitaraman et al.}

\begin{abstract}
Greenhouse gas emissions from the residential sector represent a large fraction of global emissions and must be significantly curtailed to achieve ambitious climate goals. To stimulate the adoption of relevant technologies such as rooftop PV and heat pumps, governments and utilities have designed \textit{incentives} that encourage adoption of decarbonization technologies. However, studies have shown that many of these incentives and/or subsidies are inefficient since a substantial fraction of spending does not actually promote adoption. Further, these incentives and/or subsidies are not equitably distributed across socioeconomic groups. In this paper, we present a novel data-driven approach that adopts a holistic, emissions-based and city-scale perspective on decarbonization.  
We propose an optimization model that dynamically allocates a total incentive budget to households to directly maximize the resultant \textit{carbon emissions reduction} -- this is in contrast to prior work, which focuses on metrics such as the number of new installations.  We leverage techniques for the multi-armed bandits problem to estimate \textit{human factors}, such as a household's willingness to adopt new technologies given a certain incentive.
We apply our proposed dynamic incentive framework to a city in the Northeast U.S., using real household energy data, grid carbon intensity data, and future price scenarios.  We show that our learning-based technique significantly outperforms an example of status quo incentives offered by a state and utility, achieving up to 32.23\% higher carbon reductions. Additionally, our incentive allocation approach is able to achieve significant carbon reduction even in a broad set of environments, with varying values for carbon intensity of the grid and gas and electric prices. Finally, we show that our framework can accommodate \textit{equity-aware} constraints to preserve an equitable allocation of incentives across socioeconomic groups while achieving 78.84\% of the carbon reductions of the optimal solution on average.
\end{abstract}

\maketitle

\vspace{-0.25cm}
\section{Introduction}
\vspace{-0.05cm}
\label{sec:intro}
Many countries, regions, and cities in the world have set ambitious goals for decarbonization, e.g., \textit{net zero CO$_2$ emissions} by 2035 or 2050~\cite{NetZero:2023}. 
Any realistic climate scenarios to net zero by 2050 
require deep reductions in both the energy usage and carbon emissions of buildings, industry, and transportation as well as improvements in energy efficiency throughout the economy~\cite{NGFS:2023}. 
Residential buildings accounted for 22\% of global energy consumption in 2020~\cite{ABC:2020:Global}, where a significant portion is due to carbon-intensive energy uses such as the direct combustion of fossil fuels such as natural gas for residential heating~\cite{IEA:2022:ByFuel}.
Furthermore, many fuel-based appliances and heating, ventilation, and cooling (HVAC) systems are less energy efficient than their electric counterparts~\cite{Nadel:2016:Comparative}. 
Therefore, any reductions in the carbon emissions from the residential sector can contribute significantly to the reduction of global emissions. 

One of the most promising approaches to deeply decarbonize the residential sector is to \textit{electrify} energy demand, i.e., eliminate appliances and HVAC systems that directly emit carbon, and simultaneously \emph{deploy} distributed energy resources (DER) such as rooftop solar and battery storage to ensure a steady supply of carbon-free electricity~\cite{Sitaraman:23}. 
Improvements in energy efficiency for many household appliances, such as heat pumps, and decreases in costs have made these technologies a safe and affordable option for a wide range of geographical regions and climate zones~\cite{CADMUS:2022:ASHP}. 
For instance, the cost of rooftop solar decreased by 65\% between 2010 and 2020, and combination solar and storage system level costs have decreased by 25\% since 2016~\cite{Feldman:2021:SolarCost}. 

While these decreases in cost, improvements in energy efficiency, and climate-friendly policies have enabled many transitions, these benefits are still not affordable for people across all climate regions and socioeconomic backgrounds.

There are existing incentives, in the form of rebates or tax credits, to promote electrification and energy efficiency improvement. 
For example, in the United States, a federal solar Investment Tax Credit (ITC) offers a 30\% tax credit for rooftop or community solar systems installed from 2022 through 2032~\cite{EnergyGov:2023:ITC}. There are similar state-level incentives as well. For example, Massachusetts offers 15\% of the installation costs as a tax credit (\$1000 max) with a sales tax exemption (6.25\%), New York offers 25\% of the cost as a tax credit (\$5000 max), and New Jersey offers a generous net-metering program~\cite{Forbes:2023:StateSolar}. There are additional federal incentives of up to 30\% of the costs (\$3200 max) for energy-efficient appliance upgrades such as heat pumps, water heaters, and electric panels~\cite{EnergyGov:2022:Star}. 
However, these incentives are standardized at the national or state level, and do not explicitly account for variations in energy burden for communities and their socioeconomic backgrounds, and often have prohibitive costs, e.g., the need to own the system outright. 
Such limitations prevent many households from leveraging the incentives available to them, and therefore exacerbate existing inequities.
As a result, the current incentives neither maximize the reductions in carbon emissions per dollar spent, nor do they distribute incentive budgets in an equitable manner.

To tackle this problem, we present a novel approach that adopts a holistic, city-scale framework for incentive allocation in residential sector decarbonization.  
We first present an optimization model that maximizes the \textit{reduction in carbon emissions} instead of simply maximizing the number of installations, given a total incentive budget. 
Our initial model determines the amount of incentive to offer to each household given a combination of information about their responsiveness to incentives and resultant carbon emissions.  
However, in the real world, each household's reaction to incentives is not known to decision makers.
To overcome this challenge, we leverage techniques from multi-armed bandits to estimate \textit{human factors}, such as a household's willingness to adopt new technologies given a certain incentive.
We then leverage these estimates to optimize the decarbonization plan holistically at a city-scale. 
Finally, we extend our dynamic incentive allocation approach to incorporate equity considerations by ensuring the desired allocation of incentive budget to specific socioeconomic groups.

Our proposed approach occupies a unique space within the broader literature that has explored various aspects of residential sector decarbonization. 
Many prior studies investigate the efficacy of various incentive mechanisms such as rebates and tax credits~\cite{Matisoff2017},  highlight the racial and income disparities in their adoption~\cite{Crago2017, Kearns2022}, or understand the mechanics of incentive acceptance and devise better incentive structures. 
However, most of these studies optimize incentive design for the rate of adoption, which does not always lead to maximum reductions in carbon emissions (details in \autoref{sec:relwork}). 
Finally, our work also significantly extends prior work on residential heating decarbonization that either estimates bounds on reductions in carbon emissions~\cite{Sitaraman:23, WamburuGrazierIrwinCragoShenoy:2022} or takes the central perspective of a utility in dictating which homes to transition~\cite{Lechowicz:23HP}. 

In designing our \textit{dynamic incentive allocation} approach, we make the following contributions. 

\begin{enumerate}[leftmargin=*, itemsep=0.05cm]
    \item We formalize the task of incentive allocation as an optimization problem with the objective of maximizing carbon emissions reductions, rather than adoption. Given knowledge of each household's responsiveness to incentives, our approach yields the maximum reduction in carbon emissions for a given incentive budget. 
    \item We incorporate a learning-based approach to estimate \emph{human factors} in incentive allocation, specifically household responsiveness to incentives, which is  not typically known in real-world scenarios. Our approach involves a two-stage process: in the surveying stage, we adopt a contextual offline multi-armed bandit model \cite{rashidinejad2022bridging,jin2021pessimism,li2022pessimism} with unique context, arm, and reward designs specifically tailored for deep decarbonization. For this stage, we propose a contextual lower confidence-bound method that is robust when faced with limited survey data. Then, in the offering stage, we leverage the learned best-expected carbon reduction per monetary incentive alongside the aforementioned incentive optimization technique to identify good incentives at a city-scale. Our approach can incorporate arbitrary cost models that households might use to inform incentive decisions.  We also extend our model to incorporate equity considerations in its incentive allocation approach. 
    \item We use real energy usage data from 3000+ homes in a small city in the northeastern United States to evaluate our approach. Our experimental results demonstrate that our proposed approach achieves up to 32.23\% higher reductions in carbon emissions compared to status quo incentives offered by e.g., utility companies and governments. Further, our approach achieves an average of 78.84\% of the optimal carbon reduction (with knowledge of incentive responsiveness) even under equity constraints, demonstrating that it is possible to achieve both carbon reduction and socioeconomic equity. Additionally, our incentive allocation approach is able to achieve significant carbon reduction even in a broad set of environments, with varying values for carbon intensity of the grid and gas and electric prices.    
\end{enumerate}

\vspace{-0.25cm}
\section{Background}
\label{sec:bkgd}
In this section, we present some background and discuss challenges for our incentive optimization design approach.

\smallskip
\noindent\textbf{Natural gas HVAC and appliances. \ }

Natural gas-based HVAC systems use the combustion of natural gas to generate heat for space heating. 
In other gas-powered appliances (e.g., stoves, water heaters) the generated heat is directly used to cook food or heat water. 
Since natural gas is cheaper than electricity in most locations~\cite{DOE:2023:EnergyCost}, the operating cost of natural gas-based HVAC and appliances is often lower. However, they require significant upfront investment and are hazardous due to potential methane leaks, carbon monoxide leaks, and health concerns~\cite{Gruenwald:22}.
Net-zero targets set by various jurisdictions, the availability of superior electric alternatives, and the increasing cost of maintaining a gas distribution network with a dwindling customer base suggest that societies will eventually transition away from gas-powered HVAC systems and appliances.  

\noindent\textbf{Replacement appliances for electrification. \ }

Electric heat pumps offer an energy-efficient alternative to gas-based furnaces for heating during the winter and cooling during the summer. They leverage heat transfer instead of heat generation and can deliver up to 7 times more heat energy than the electrical energy they consume~\cite{Johnson:2013:Measured, Nadel:2016:Comparative}.  
While electric heat pumps have been around for decades, their heating performance in cold climates, such as in the northeast United States, has been poor. However, recent technological improvements have made their deployment viable even in colder climates. 
For other appliances, such as water heater and stoves, electric alternatives are efficient and have lower purchase prices.  However, due to diversity in performance (e.g., radiant vs. induction stoves), adoption of these alternatives must overcome hurdles of public perception and existing expectations of electric appliance performance.

\noindent\textbf{Electric grid. \ }

Electric power is traditionally generated at large power plants, often powered by fossil fuels such as natural gas, oil, and coal~\cite{US:2023:Electricity}. 
The generated electricity travels long distances through transmission networks and is distributed to individual homes using a distribution network. This traditional grid poses two key challenges to electrification for decarbonization. First, added demand on the grid due to electrification can trigger expensive upgrades in the electricity distribution and, in some cases, transmission networks~\cite{ACEEE:2023, GreenAlliance:2023}. Second, the electric grid is still evolving to handle large-scale intermittent and inverter-based resources, such as solar and wind~\cite{Shair:2021:Power}. Therefore, the electric grid may not be entirely carbon-free in the near future~\cite{EIA:2022:Outlook}, preventing full decarbonization of the residential sector despite electrification. A holistic decarbonization of the residential sector should be cognizant of these challenges and target cost-effective deep decarbonization.

\noindent\textbf{Residential solar and battery storage. \ }

Electrification of the residential sector coupled with co-located solar and battery storage systems offers a cost-effective solution for deep decarbonization~\cite{Sitaraman:23}. Solar and battery storage systems are cost-effective as they do not require fuel and have a long life (25 years for solar and 5-15 years for batteries). 
Residential solar projects can be installed without battery storage to reduce cost, although avoiding curtailment would require demand to be larger than generation at all times, or the ability to backfeed excess energy to the grid.  However, many states in the United States do not have attractive net-metering programs, and some are proposing high grid access charges~\cite{NCCETC:2023:FiftyStates}. Furthermore, a solar system alone may still require grid upgrades, as it cannot reduce the peak demand for electric heating (e.g., at night or during inclement weather). On the other hand, a solar system with appropriately sized storage can mitigate added electricity demand, allowing a deep decarbonization. 

\noindent\textbf{The social cost of carbon. }The Social Cost of Carbon (SCC) assigns a dollar value to the future net social, economic, and physical impacts of 1 metric ton of carbon~\cite{epa_scc_report}. In the United States, this value is estimated by the Environmental Protection Agency (EPA) through projection modules for emissions, population, income, climate, damages, and discounting based on the latest developments in the relevant fields \cite{epa_scc_report}. A discounting module is used to determine future climate damages occurring during a specific year of emissions and is based on a specific dynamic discount rate calculated with the Ramsey discounting formula. Due to uncertainty regarding the starting rate, the EPA has provided calculations of the SCC based on 3 different likely near-term target rates, 1.5\%, 2\% and 2.5\%, which provide a high, medium, and low estimate for the SCC. For the purposes of our work, we use the SCC calculated with a near-term target rate of 2\%.

\section{Problem \& Preliminaries}
\vspace{-0.05cm}
\label{sec:prob}
In this section, we formulate the incentive allocation problem with the objective of finding an allocation that maximizes the reduction of carbon emissions subject to a budget constraint.
 
We present the problem assuming that each household's willingness to adopt the decarbonization plan given a certain incentive is known a priori to the incentive designer (we relax this assumption in \autoref{sec:methods}).  
We also motivate the dynamic incentive design problem by analyzing financial ``break-even'' points for the deep decarbonization retrofit package. 

\subsection{Problem formulation}
\label{sec:opt}

We first present a general optimization framework that assumes knowledge of key problem parameters in advance.

We denote the set of households in, e.g., a city by $\mathcal{H}$, and let $n = \vert \mathcal{H} \vert$.
We let $E^{\texttt{CO}_2}$ represent the carbon intensity of the electric grid, and $G^{\texttt{CO}_2}$ is a constant describing the carbon emissions due to natural gas combustion (in grams of CO\texttt{$_2$} emitted per unit of energy). 

Suppose the incentive designer has a total budget of $B$ (e.g., in USD).
An \textit{incentive allocation} $\mathbf{I}$ is a vector in $\mathbb{R}^n$, where the $h^\text{th}$ term of $\mathbf{I}$ is the monetary incentive offered to house $h \in \mathcal{H}$.  Then, the space of valid incentives is described as $\mathcal{I} \coloneqq \{ \mathbf{I} \in \mathbb{R}^n \ : \ \lVert \mathbf{I} \rVert_1 \leq B \}$.

Let $A_{(h | \mathbf{I})}$ denote the conditional probability that house $h \in \mathcal{H}$ adopts the deep decarbonization package under incentive allocation $\mathbf{I}$.  For brevity, we will refer to $A_{(h | \mathbf{I})}$ as an \textit{acceptance function} that yields this probability.  The objective of the incentive designer is to maximize the objective function subject to budget constraints. Formally, we define the offline optimization problem as
\begin{align}
    \max_{\mathbf{I} \in \mathcal{I}} \; & \;\sum_{\mathclap{h \in \mathcal{H}}}\; A_{(h | \mathbf{I})} \left[\left(g(h) G^{\texttt{CO$_2$}} + \left[ e(h) - e'(h) \right] E^{\texttt{CO$_2$}} \right)\times \text{SCC} \right], \label{eq:obj} \\
    \textnormal{s. t.,} \; & \; \lVert \mathbf{I} \rVert_1 \leq \; B. \label{eq:cons}
\end{align}

Here, $g(h)$ denotes the reduction in home gas usage. $e(h)$ and $e'(h)$ denote the yearly grid electric usage of house $h$ \textit{before} and \textit{after} the deep decarbonization package is installed, respectively.  Note that this captures any excess electricity that must be pulled from the grid beyond what solar and storage supply. The $\text{SCC}$ value is a constant that represents the Social Cost of Carbon for the given year during which this optimization problem is considered. 

In the setting where $A_{(h | \mathbf{I})} \in \{0, 1\}$ and is known for each house, we can reduce this optimization to a \textit{knapsack problem} as follows:  Let $w_h$ be the minimum value of $\mathbf{I}_h$ (i.e., the incentive for house $h$) such that $A_{(h | \mathbf{I})} = 1$. 
Let $v_h$ be carbon emissions reduction (($g(h) G^{\texttt{CO}_2} + \left[ e(h) - e'(h) \right] E^{\texttt{CO}{_2}} )\times \text{SCC} $). Then we have the following knapsack problem: $\max \sum_{h\in \mathcal{H}} x_h v_h$ s.t. $\sum_{h\in \mathcal{H}} x_h w_h\le B$, where the value and the weight of each ``item'' are $v_h$ and $w_h$ (respectively), and $x_h \in \{0, 1\}$ is a decision variable indicating whether an incentive is paid out to house $h$.

\noindent\textbf{Equity component.}
In this section, we describe the \textit{equity} constraints that we will impose on the above formulation in \autoref{sec:evalEq}.  Suppose the incentive designer would like to specify an equitable incentive distribution across $M$ groups (e.g., socioeconomic groups).  Instead of a single budget $B$, we define $M$ budgets such that each $B_m$ for $m\in [M]$ corresponds to the total budget allocated to the $m^\text{th}$ group.  Then, the constraint in \autoref{eq:cons} is replaced with $M$ constraints, such that the components of $\mathbf{I}$ corresponding to group $m$ (denoted by $\mathbf{I}_m$) must be less or equal to $B_m$.  Formally, we have
\begin{equation}
\textnormal{s. t.,} \; \; \lVert \mathbf{I}_m \rVert_1 \leq \; B_m \; \; \forall m \in [M]. \label{eq:consEquity}
\end{equation}

\noindent\textbf{Decarbonization over time.}
In the case where the incentive designer must offer incentives over the course of many years, and is only able to access a fraction of the budget $B$ each year, we describe a modification to the optimization in \autoref{eq:obj} and \autoref{eq:cons}. 
In this scenario, the incentive designer projects the value of the carbon emission reduction over a number of years $Y$. For a given year $ y\in Y$,  $v^{(y)}_h$ represents the projected carbon emission reduced for home $h$ given that the home is decarbonized during year $y$. Thus, for this scenario, for each year $y \in Y$, we use a modification of \autoref{eq:obj} as follows:
\begin{align}
    \max_{\mathbf{I^{(y)}} \in \mathcal{I}} & \sum_{\mathclap{h \in \mathcal{H}_y}}\; A_{(h | \mathbf{I^{(y)}} )} \left[ \sum_{\mathclap{t = y}}^{Y} \left( g(h) G^{\texttt{CO$_2$}} + \left[ e(h) - e'(h) \right] E^{\texttt{CO$_2$}} \right) \text{SCC}_t \right], \label{eq:objot} \\
    \textnormal{s. t.,} \; & \; \lVert \mathbf{I^{(y)}} \rVert_1 \leq \; \nicefrac{B}{Y}, \label{eq:consot}
\end{align}
where $\text{SCC}_t$ is the value of the social cost of carbon for a given year $t$, $\mathcal{H}_y$ are the homes in $\mathcal{H}$ that have not yet accepted an incentive by year and the value of 
an incentive allocation vector $\mathbf{I}_y$ is determined, where $\lVert \mathbf{I}_y \rVert_1 \leq \nicefrac{B}{Y} \;$.

\noindent\textbf{Equity constraints on decarbonization over time.}
In this section, we impose equity constraints on the optimization framework presented in above with \sref{Equations}{eq:objot} and \ref{eq:consot}. We describe two different equity constraints, which we refer to as \textit{Strict Equitable Allocation} and \textit{Relaxed Equitable Allocation}. \textit{Strict Equitable Allocation} ensures that the yearly budget is split equitably across $M$ groups. Thus, the constraint on the yearly budget is similar to the constraint given in \autoref{eq:consEquity}. Specifically, it is 
\begin{equation}
\; \; \lVert \mathbf{I}_m^{(y)} \rVert_1 \leq \; B^{(y)}_m \; \; \forall m \in [M], \label{eq:consStrictEquity}
\end{equation} 

where $B^{(y)}_m$ is the portion of the budget from year $y$ allocated to group m, and $\mathbf{I}_m^{(y)}$ is the incentives allocated for group $m$ during year $y$. 
\textit{Relaxed Equitable Allocation} relaxes the constraint that a portion of the yearly budget $B^{(y)}$ must be allocated to a group $m \in M$. Rather, it splits the overall budget $B$ across the $M$ groups just as in \autoref{eq:consEquity}. The incentive designer is still only able to use a total of $B/Y$ per year, but the maximum that can be spent on each group $m$ over the total number of years $Y$ is $B_m$. That is, 
\begin{equation}
\; \sum_{\mathclap{y = 1}}^{Y}\; \lVert \mathbf{I}_m^{(y)} \rVert_1 \leq \; B_m \; \; \forall m \in [M]. \label{eq:consRelaxedEquity}
\end{equation} 

Where $Y$ is the total number of years, $B_m$ is the total budget that group $m$ can use, and $\mathbf{I}_m^{(y)}$ is the incentives allocated to group $m$ during year $y$. 
In the offline formulation, it is assumed that the acceptance functions $A_{(h | \mathbf{I})}$ are known to the incentive designer.  This is an unreasonable assumption in practice, and is addressed in \autoref{sec:methods}.  
We next discuss a model from the literature on cost/benefit analysis that will attempts to quantify whether a given household will ``opt-in'' to a decarbonization plan for a certain incentive.  We use this model to motivate the need for external incentives.

\subsection{Modeling likelihood of adoption}
\label{sec:costmodel}

Prior work on decarbonization technologies, such as rooftop PV and air source heat pumps, finds that the likelihood of a household adopting technology is a function of the return on investment (RoI)~\cite{Bjrnstad2012, Zander2019, Barnes2020, Kearns2022, Sher2022, PobleteCazenave2023}, where returns refer to the cost savings after adopting the decarbonization technology.
Since our optimization formulation requires a definition of the acceptance function, $A_{(h | \mathbf{I})}$, for each house $h \in \mathcal{H}$, we leverage prior work to devise a cost model that may inform a hypothetical household's decision to adopt a decarbonization package when offered with a certain incentive. 
To motivate the need for incentives, we analyze the financial ``break-even''  points for the households in our data set (details in \autoref{sec:datasets}), as without them, the investment outweighs the expected returns for a large fraction of households, prohibiting adoption.

\noindent\textbf{Net Present Value (NPV).} A standard method for break-even analysis in the literature is termed {\em Net Present Value} (NPV), which recognizes that a monetary benefit in the future is worth less at present due to inflation, interest rates, and other factors \cite{Lee:2017}. 
This effect is captured using {\em discount rate}, a factor specifying how much future returns are worth at present. 
The {\em net benefit} of the conversion is the NPV of the future benefits minus the conversion's cost. 
These benefits include the money saved on annual gas and electric bills during the recovery period $T$ (in years). For a given house $h$, the net benefit can be calculated as
\[
\textnormal{NetBenefit}(h) \coloneqq \sum_{t=0}^{T} \frac{(c_b( g(h), e(h)) - c_b(e'(h))}{(1 + \textnormal{discount\_rate})^t} - c_e(h).
\]

Where $c_b(\cdot)$ indicates the annual cost in dollars at the beginning of the time window considered, i.e, $c_b(e(h), g(h))$ is the annual cost of the gas and electricity of home $h$ before implementing a decarbonization package, and  $c_e(\cdot)$ indicates the annual cost of electricity after implementing the decarbonization package. If the NetBenefit quantity is \textit{positive}, the future benefits outweigh the cost, and vice versa.  The impact of an incentive given to a household, $\mathbf{I}_h$, can be captured by substituting $- c_e(h)$ with $+ \mathbf{I}_h - c_e(h)$.
We set $A_{(h | \mathbf{I})} =1$ if If $\mathbf{I}_h$ leads to a positive NetBenefit. 

A reasonable choice for the recovery period $T$ should be less than or equal to the expected lifespan of the decarbonized energy system (i.e., how long it will last without replacing any components). For instance, residential lithium-ion battery systems typically have a lifespan between 5-15 years~\cite{Solar_Calculator_2024} -- thus, in our analysis, we set $T$ to denote a ``break-even'' threshold between 5-15 years.

NetBenefit values are highly dependent on the discount rate values, which are determined by inflation, electricity price growth, and prevailing interest rates. Thus, the economic situation of a homeowner's country will contribute greatly to their household's ``break-even'' point. In implementing our NetBenefit model, we consider two discount rate scenarios.  
In the first scenario, we assume moderate inflation and electricity growth rates. In the second scenario, we assume high inflation and a high electricity growth rate. 
In the U.S., estimates for discount rates are based on historical data for federal funds rates, inflation rates, and increases in electric prices. The moderate electricity price growth rate is based off of a historical regional electricity growth rate of 10\% \cite{jkinney@repub.com_2022}, the federal funds rate is assumed to be 5\%, and the moderate inflation rate, which is the average inflation rate across 2013 to 2019, is assumed to be about 1.5\% \cite{USInflationCalculator}. For the second discount rate, the inflation rate is assumed to be 8\%, based on the average inflation rate in 2022, and the electricity price growth rate is assumed to be 11\% based on the average electricity price growth rate in 2022 in the United States.

\begin{figure}[t]
    \centering
    \includegraphics[width=0.78\linewidth]{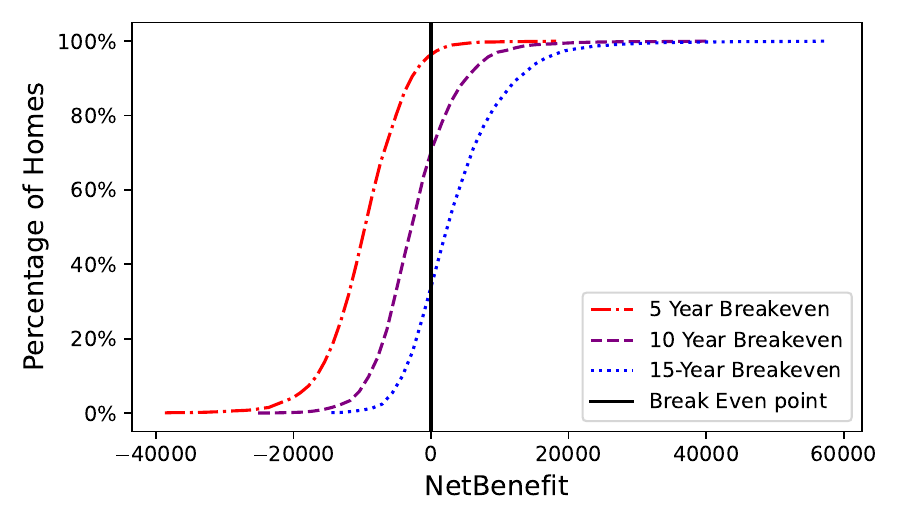}
    \vspace{-0.5cm}
    \caption{A representative break-even point illustration under the NetBenefit model with a 5, 10 and 15 year payback period; 96\%  of houses fail to break even with the installation costs of decarbonization within 5 year payback period, 70\% of houses fail to break even within 10 year payback period, and 33\% of houses fail to break even within a 15 year payback period.} 
    \label{fig:break_even}
    \vspace{-0.5cm}
\end{figure}

In \autoref{fig:break_even}, we observe that for a NetBenefit cost model under the moderate discount rate scenario, 96\% of homes in our dataset (those on the left-hand side of the black line) will not ``break-even'' within five years of adopting a decarbonization package, 70\% will not break even within 10 years, and 33\% will not break even within 15 years (data and methodology explained in \autoref{sec:expsetup}).
This implies that most households will not be financially motivated to decarbonize under the status quo, motivating the need for a strategic incentive allocation to households that reduces the upfront financial burden of a deep decarbonization package.

While the cost model discussed here is grounded in the relevant literature, it is also known~\cite{Blumsack2012, Zhao2012, Bauner2015, Snape2015} that the true behavior of homeowners when presented with decarbonization options and incentives is not easily predictable by simple models assuming ``rational actors''.  Thus, in the next section, we describe our approach to essentially \textit{estimate} $A_{(h | \mathbf{I})}$ for each house $h \in \mathcal{H}$, rather than attempting to model them directly.  We use the cost models discussed above as an approximation of true behavior throughout our experiments.

\vspace{-0.2cm}
\section{Survey Learning}
\label{sec:methods}
In this section, we describe our
approach to generalize the solution design presented in the previous section (i.e., with knowledge of each household's incentive response) to the more practical case of unknown acceptance functions.  

We propose a \textit{two stage approach} where in the first stage, 
we leverage algorithmic foundations from \textit{offline multi-armed bandits} to survey a subset of households and estimate acceptance thresholds. In the second stage, we then use these estimated thresholds in the solution design described above to choose households throughout the city that should be incentivized.

\subsection{The surveying stage}\label{sec:survey}
A traditional survey approach, as would be typical conducted when understanding community attitudes towards decarbonization technologies for example, may present households with different options and collect responses.  Such an approach is unsuitable for the application we consider.  If an incentive designer presents households with several ``tiers'' of financial incentives and asks them for the minimum incentive they would accept, it will not elicit a truthful response. This motivates a ``limited information'' setting, described below.

Instead of presenting each household with several options, our proposed survey design offers each household a decarbonization package and a specific monetary incentive.  Each household can then respond to this offer with a simple ``accept-reject'' response.  The incentive designer's role is to choose a decarbonization package and incentive to offer and can observe the household's accept-reject response to that specific offer.  This exactly mirrors the concept of \textit{bandit feedback}, and by nature, is more likely to result in truthful responses because survey respondents cannot e.g., select the highest offered incentive and claim this as the minimum they would accept.

\noindent\textbf{The contextual offline bandit model. \ }
We frame the surveying stage described above as a \textit{contextual offline bandits problem}
~\cite{li2022pessimism,rashidinejad2022bridging,jin2021pessimism}, where $N$ survey responses are collected from different households in parallel.  
In the general contextual offline bandits problem, the learner must contend with contexts that influence the reward associated with each of the $K$ arms (i.e., actions).  

The learner's objective is to determine the optimal arm for each context when given $N$ examples to learn from, where each example consists of an arm, a context, and the resulting reward. 

In our setting, the \textit{context} of each household $h$ refers to reasonably accessible information about each household in the city that can be obtained by the incentive designer.  
The offers that the incentive designer can send to households are captured by the $K$ arms of the contextual bandit algorithm, each of which corresponds to a unique (decarbonization package, incentive) pair.  
Finally, the \textit{reward} of each offer is defined as the carbon reduction (if the offer is accepted) divided by the amount of incentive.  We describe each of these in detail below.

To construct a ``data set'' of examples (i.e., survey responses) for the learner, we select a subset of $N$ households to receive surveys, where $N < \vert \mathcal{H} \vert$.  As mentioned above, each household $h$ has a context that is recorded.  Each survey sent out includes an offer that is uniformly randomly chosen from the $K$ arms.\footnote{Note that if the incentive designer has some prior information about which ``arms'' may be the optimal incentive scheme, they can introduce bias in this process and generate survey offers that provide more coverage of the likely-optimal arms.}

\noindent\textbf{Context design. \ }
In our setting, inspired by the cost model described in \autoref{sec:costmodel}, we use three variables to describe the \textit{context} of each household, namely, \textit{median income, yearly gas usage, and yearly electric usage}.  Each of these variables can be obtained by the incentive designer from, e.g., publicly available census data or utility data.  To discretize the possible contexts, we split each variable by quantiles into $5$ groups, such that each household falls into one of $C=5^3=125$ possible contexts.

We note that including variables such as median income can also help facilitate \textit{equity-aware} optimization of incentives as in~\cite{Lechowicz:23HP, WamburuGrazierIrwinCragoShenoy:2022}, by allowing the incentive designer to target surveys and eventual incentives towards neighborhoods which have historically been disadvantaged by existing energy systems~\cite{Sunter:19, Baker:21}.

\smallskip \noindent\textbf{Arm descriptions.}
We design $K$ arms, where each arm $k\in[K]$ corresponds to a simultaneous choice of decarbonization package $D_k\in \mathcal{D}$ (e.g., full appliance replacement vs. just heat pump installation) and monetary incentive $I_k \in \mathbb{R}_{+}$.  We discretize the range of monetary incentives into $5$ tiers based on quantiles of the actual payoff thresholds computed using the cost model mentioned above.  This is in contrast to the full information setting, where the cost models prescribe an exact amount of money at which the household should accept an incentive.

\smallskip \noindent\textbf{Reward formulation.}

For the $j^{\text{th}}$ survey response from $N$ total responses, we say that the context of the household is given by $c_j$ and the offer is given by $k_j \in [K]$, with associated monetary incentive $I_{k_j}$ and decarbonization package $D_{k_j}$.  The feedback received from the household is encoded as a \textit{reward} $r_j$ as follows:

If the household rejects the offered incentive, the realized reward is $r_j=0$.  If the household accepts the offered decarbonization package $D_{k_j}$ and monetary incentive $I_{k_j}$, the realized reward $r_j$ is computed as the ratio of the carbon reduction to the monetary incentive, i.e., $r_j=R_j/I_{k_j}$, where $R_j= R(D_{k_j},c_j)$ represents the \textit{estimated yearly carbon reduction} based on incentive designer's data for all households in the context $c_j$ who adopt decarbonization package $D_{k_j}$. To this end, we denote the offline survey data set as $\mathcal{D}_{\text{off}} = \{(c_j, k_j, r_j)\}_{j=1}^N$.

\subsection{The offering stage} \label{sec:offering}
In the second phase, the incentive designer offers decarbonization packages and incentives to each household in the city that exactly correspond to the best-expected payoff (in terms of dollars per carbon reduction) learned from the survey results.  To learn the best incentives, we leverage algorithmic foundations from the offline multi-armed bandits literature, described below.

\smallskip \noindent\textbf{The Contextual Lower Confidence Bound algorithm.}

The goal of using a contextual bandits algorithm is to learn the expected rewards of each arm, with the specific goal of identifying an ``estimated optimal arm'' (consisting of a decarbonization package and monetary incentive) for each household context in the community. We adopt the Contextual Lower Confidence Bound (CLCB) algorithm, which operates as \( C \) independent LCB algorithms~\cite{li2022pessimism}, each corresponding to a context \( c \in [C] \). Given the offline survey data set \( \mathcal{D}_{\text{off}} = \{(c_j, k_j, r_j)\}_{j=1}^N \), CLCB tracks both the number of times each arm \( k \) is pulled within a given context \( c \), denoted by \( T_{k,c} = \sum_{j=1}^N \mathbb{I}\{c_{j} = c, k_{j} = k\} \), and the empirical mean reward \( \hat{\mu}_{k,c} = \sum_{j=1}^N \mathbb{I}\{c_j = c, k_j = k\} r_j / T_{k,c} \) if $T_{k,c}>0$ and $\hat{\mu}_{k,c}=0$ otherwise.
The LCB value for each arm \( k \in [K] \) and context \( c \in [C] \) is then computed as:
\begin{equation}\label{eq:LCB}
    \underline{\mu}_{k, c} = \max \left\{\hat{\mu}_{k,c} - \alpha \sqrt{\frac{\log N}{T_{k,c}}},0\right\},
\end{equation}
where \( \hat{\mu}_{k,c} \) represents the predicted reward (e.g., carbon reduction per dollar of incentive), and \( \alpha \sqrt{\frac{\log N}{T_{k,c}}} \) reflects the uncertainty in the reward estimate, with \( \alpha > 0 \) controlling the penalty applied to under-explored arms.
Since the survey data is pre-collected, with no possibility of further exploration, CLCB emphasizes well-explored arms while penalizing those with limited responses by reducing their predicted reward. This conservative approach enhances robustness, especially when working with limited or unevenly distributed survey data.

To quantify the number of offline data needed to find out good arm $\hat{k}(c)$ that is as close to the optimal arm $k^*(c)$ for each context $c$, we give the following analysis to bound the number of survey responses $N$. Based on \cite{rashidinejad2022bridging}, we assume an offline dataset $\{(c_j, k_j, r_j)\}_{j=1}^N$ is i.i.d. generated with context $c_j$ and arm $k_j$ following distribution $\mathbb{D}$. Denote data coverage $C^* = \max_{c} \frac{1}{\mathbb{D}(c, k^*(c))}$, where $k^*(c)$ is the optimal arm for context $c$. In the survey stage, an upper bound of $O\left(\frac{C^* \log (N)}{\epsilon^2}\right)$ on sample size guarantees, with probability $1 - O(\frac{1}{N})$, that the LCB algorithm finds an arm (decarbonization package and incentive) within $\epsilon$ of optimal $k^*(c)$ for any $ c $. This theoretical guarantee comes from the following derivation:
\begin{align*}
    \mu_{k^*(c),c}-\mu_{\hat{k}(c),c}&\overset{(a)}{=}(\mu_{k^*(c),c}-\underline{\mu}_{k^*(c),c}) + (\underline{\mu}_{k^*(c),c} - \underline{\mu}_{\hat{k}(c),c}) + (\underline{\mu}_{\hat{k}(c),c}- \mu_{\hat{k}(c), c})\\
    &\overset{(b)}{\le} \mu_{k^*(c),c}-\underline{\mu}_{k^*(c),c}\\
    &\overset{(c)}{\le} \sqrt{\frac{2\log N}{T_{k^*(c),c}}} 
    \overset{(d)}{\le} \sqrt{\frac{4\log N}{N \cdot \mathbb{D}(c, k^*(c))}} 
    \overset{(e)}{\le} \sqrt{\frac{4C^* \log N}{N}} 
    \overset{(f)}{\le} \epsilon
\end{align*}
where inequality (a) is due to adding and minus terms $\underline{\mu}_{k^*(c),c}$, $ \underline{\mu}_{\hat{k}(c),c}$, inequality (b) is because the second and third terms are non-positive by inequality~\ref{eq:LCB} and $\underline{\mu}_{k,c}$ being a high probability lower bound of $\mu_{k,c}$ according to Hoeffding's inequality when setting $\alpha=\nicefrac{1}{\sqrt{2}}$, inequality (c) is due to the same reason that $\underline{\mu}_{k,c}$ being a high probability lower bound, inequality (d) holds due to multiplicative Chernoff bound when $N\ge 8C^*\log N \ge 8 \log N/\mathbb{D}(c, k^*(c))$, inequality (f) holds due to $N= \nicefrac{4C^*\log N}{\epsilon^2}$.

\smallskip
\noindent\textbf{Offer construction. \ }
Given estimates of the best incentives provided by the LCB technique, the \textit{offering stage} proceeds as follows:
Each household $h$ with context $c_h$ is presented the decarbonization package $D_{k_h}$ and the incentive $I_{k_h}$, where $k_h=\arg\max_{k\in[K]}\underline{\mu}_{k,c_h}$ is the optimal offer learned during the survey phase. 
Once this proposal is sent, house $h$ responds with a decision, i.e., whether they accept $(D_{k_h},I_{k_h})$ or not (non-response is interpreted as a rejection). We denote $\mathcal{H}_a \subseteq \mathcal{H}$ as the houses that accept this proposed incentive.\footnote{Since the incentives are discretized, it is sometimes the case that the lowest tier of incentives is nominal, meaning that the survey estimates predict that home $h$ does not need an incentive to accept a decarbonization package. In the case that $I_{k_h}$ is this lowest tier, we assume that home $h$ is offered the decarbonization package resulting in the highest possible carbon reduction rather than $D_{k_h}$.} 

Given $\mathcal{H}_a$, the incentive designer sets a budget $B$ that they are willing to spend on the current set of incentives, where $B \leq \sum_{h \in \mathcal{H}_a} I_{k_h} $.  Under this budget constraint, they compute a final set of decarbonized houses $H^*$ by solving a knapsack problem over houses $\mathcal{H}_a$, i.e., $H^*=\arg\max_{H\subseteq \mathcal{H}_a: \sum_{h\in H}I_{k_h}\le B}\sum_{h\in H}v_h$, where $v_{h}$ is the (known) carbon reduction of house $h$, according to the incentive designer's data.  Note that by solving this knapsack problem, the incentive designer can exclude those houses that achieve small carbon reductions relative to the expenditure required to incentivize them.
Furthermore, to accommodate equity constraints, we can change the budget constraint $B$ to $M$ constraints as in \autoref{eq:consEquity} and solve a new optimization problem over $\mathcal{H}_a$. 
The final incentive expenditure is $\mathbf{I}_h=I_{k_h}$ if $h\in H^*$ and $\mathbf{I}_h=0$ otherwise.

\vspace{-0.15cm}
\section{Experimental Setup}
\label{sec:expsetup}
In this section, we describe our experimental setup, including energy usage data sets,  carbon intensity data, census and property data, cost models, and algorithm implementations.

\begin{figure*}[t]
    \centering
    \includegraphics[width=\linewidth]{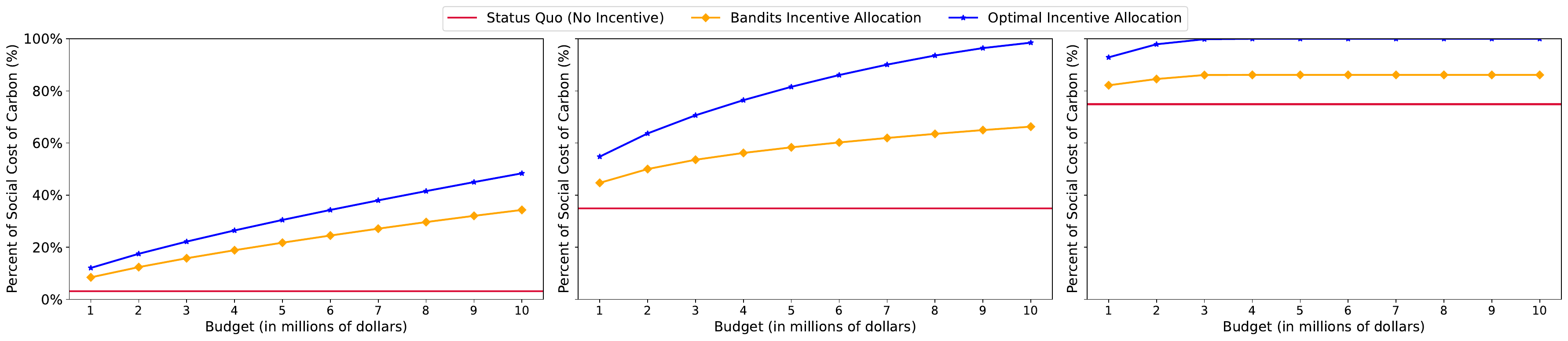}
    \vspace{-0.45cm}
    \caption{Considering the NetBenefit cost model with a \emph{modest} discount rate of 5\% for payback periods of (a) 5 years, (b) 10 years and (c) 15 years. Our learning-based incentive model achieves an average of 17.71\% higher carbon reduction than the status quo across all budgets and payback periods.} \label{fig:costmodel1nonEquity}
    \vspace{-1em}
\end{figure*}

\begin{figure*}[t]
    \c
    entering
    \includegraphics[width=\linewidth]{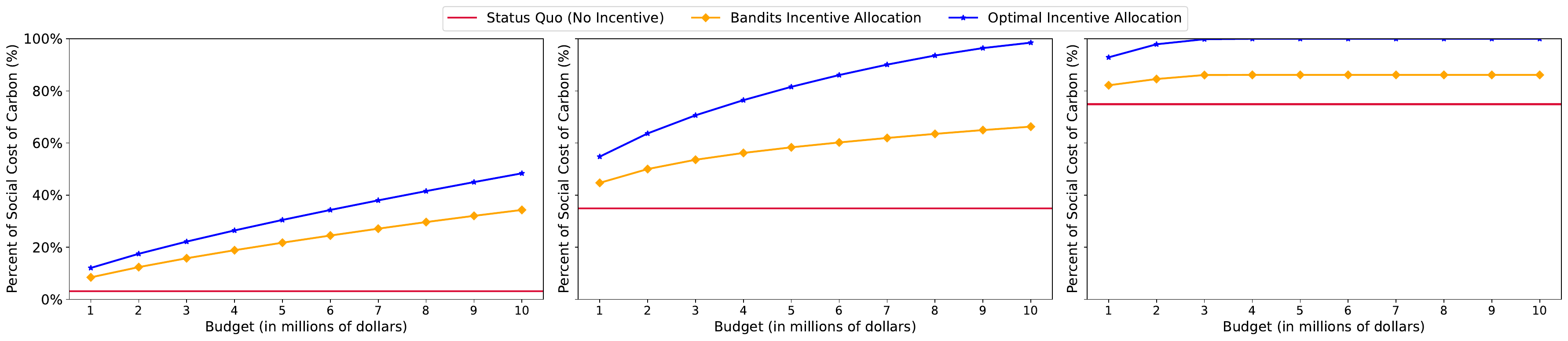}
    \vspace{-0.45cm}
    \caption{Considering the NetBenefit Cost Model with a discount rate of 2\% for payback periods of (a) 5 years, (b) 10 years and (c) 15 years. Our learning-based incentive model average of 15.22\% higher carbon reduction than the status quo across all budgets and payback periods} \label{fig:costmodel2nonEquity}
    \vspace{-1em}
\end{figure*}

\subsection{Data sets} 
\label{sec:datasets}

\noindent\textbf{Household energy use.}
We use natural gas and electric consumption data for 3,168 individual housing units in a small city in the northeastern U.S., which spans all of 2020 and provides a representative snapshot of usage across seasons.  
Electricity consumption data is provided at a 5-minute granularity, while gas consumption is at a one-hour granularity.

\noindent\textbf{Grid carbon intensity data.}
We obtain historical grid carbon intensity data via Electricity Maps~\cite{electricity-map} for five grid regions in the U.S., which include Bonneville Power Administration (\texttt{BPAT}), California ISO (\texttt{CAISO}), ISO New England (\texttt{ISO-NE}), Pennsylvania-Jersey-Maryland (\texttt{PJM}) interconnection, Pacificorp East (\texttt{PACE}) and South Carolina Public Service Authority (\texttt{SC}). 

This data provides each grid's hourly \textit{average carbon intensity}, expressed in grams of CO$_2$ equivalent per kilowatt-hour (gCO$_2$eq/kWh).  These regions represent a variety of grid \textit{generation mix} scenarios, e.g., \texttt{BPAT} represents a highly clean grid powered by hydropower, whereas \texttt{SC} represents a dirty grid fed mostly by coal and gas.  Other regions fall in the middle, and their ordering above represents their order in terms of yearly average carbon intensity.

\noindent\textbf{Census data.}
We use U.S. Census Bureau reports to get the basic socioeconomic characteristics of the households in our dataset~\cite{Census:2020}.  
We classify neighborhoods as low, medium, and high income using their census-reported median income, an approach used by prior work on equitable residential decarbonization~\cite{WamburuGrazierIrwinCragoShenoy:2022, Lechowicz:23HP}.

\noindent\textbf{Solar potential data.}
For each household in the data, we use the Solar-TK toolkit~\cite{SolarTK} to obtain per-building solar potential. The toolkit estimates the output of a solar installation (with given parameters) at the given location over the course of an entire year.

\subsection{Cost modeling and implementations}

\noindent\textbf{Cost modeling.} 
We estimate the cost and sizing of equipment needed for each household as follows.  
A solar installation is sized based on the energy demand for the household and the available roof area. The battery storage is sized to store the maximum surplus solar, i.e., solar generation minus daytime demand. The solar and storage system installation costs are assumed to be \$2,002 USD per kW and \$1,047 USD per kWh, respectively. These costs are based on current average costs \textit{after federal and state tax credits} in the area under study~\cite{SolarStorage:2024:Prices} and include materials and labor costs.   
  
An air-source heat pump retrofit is sized for each household by converting the current gas usage to an equivalent thermal output, as in~\cite{Lechowicz:23HP}. 
We estimate the electricity demand in kWh needed to generate the same thermal output, using an ambient temperature-dependent model for heat pump coefficient of performance (COP), as in prior work~\cite{KellyCockroft:2011}. 
To estimate the installation cost, we find a household whose usage is approximately the median amongst households that use natural gas for heating.  For this household, we set the installation cost for a high-efficiency heat pump system to the industry average of \$5,250 USD~\cite{Davis2023}. We estimate the installation cost for other households by scaling the benchmark cost proportional to the gas usage, e.g., the heat pump for a household with gas usage $3\times$ the median will cost \$15,750 USD.

We use natural gas usage in \textit{summer months} as a proxy for the presence of non-heating related natural gas appliances, such as a water heater, based on prior work~\cite{Lechowicz:23HP}. The replacement cost for a 120V heat pump water heater is set to the industry average after rebates, \$1575 USD~\cite{energystarwaterheater}.
We consider two decarbonization packages: a ``just heat pump'' option, where only the portion of natural gas used for home heating is replaced by an air-source heat pump, solar PV installation and battery, and a ``full appliance replacement'' option, where natural gas usage is entirely eliminated and all gas-based heating and appliances are replaced with electric appliances, air-source heat pumps as well as a solar PV installation and battery.  

In our experiments, we set a natural gas price of \$1.160/CC and an electricity price of \$0.14072/kWh, unless stated otherwise. These values incorporate the cost of maintaining a gas and electric meter and represent the actual prices in the city under analysis~\cite{HGE:2023:Prices}. 

\noindent\textbf{Algorithmic implementations.} 
We implement two incentive allocation schemes in Python as follows.  We use numerical optimization methods provided by SciPy~\cite{SciPy} to compute the optimal solution to the problem summarized in \autoref{sec:opt}, leveraging perfect knowledge of the acceptance functions.  This gives an \textit{upper bound} on the carbon reduction achievable for a given budget if, e.g., the preferences of all housing units are known exactly to the incentive designer.  For our learning-based incentive allocation, we implement the contextual LCB algorithm~\cite{auer2002nonstochastic} described in \autoref{sec:methods}.  For each cost model and payback period, we run a preliminary ``survey phase'', which surveys $N$ random housing units (unless otherwise specified, $N=1000$).  
The survey phase produces a data set that is used to estimate the optimal ``arm'' (incentive) for each of 125 contexts. These estimates are then used to offer and allocate incentives according to the description in \autoref{sec:offering}. 
When the experiment-prescribed budget is large, it is sometimes the case that the initial round of accepted incentives (i.e., $\mathcal{H}_a$, see \autoref{sec:offering}) does not take up the entire budget.  In this case, we send a single extra round of incentives.  In this extra round, the remaining households are each offered an incentive that is one ``tier'' larger than the optimal incentive learned by the LCB algorithm (recall that the incentives are discretized into 5 ``tiers'' during the survey stage, see \autoref{sec:survey}).  We then solve the knapsack problem to select between the households that accepted either round of incentives.

\section{Experimental Results}
\label{sec:eval}
In this section, we evaluate our learning-based dynamic incentive framework termed ``Bandit Incentive Allocation'' that leverages data-driven insights. 
The baseline approaches include the ``Status Quo'' which assumes no incentives beyond those already embedded in the cost models and includes homes that would adopt decarbonization technologies even without additional incentives, and ``Optimal Incentive Allocation,'' representing the approach in \autoref{sec:opt} that assumes full knowledge of each household's willingness to accept incentives. 
\begin{figure*}[t]
    \centering
    \includegraphics[width=\linewidth]{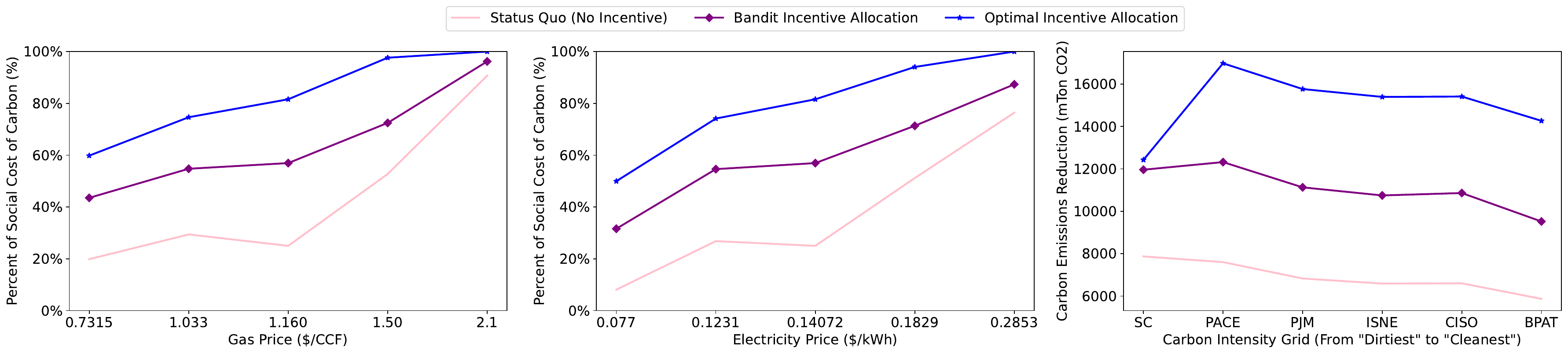}
    \vspace{-0.45cm}
    \caption{The effect of changing environments characterized by (a) gas price, (b) electricity price, and (c) carbon intensity.} \label{fig:environments}
    \vspace{-1em}
\end{figure*}

The two baseline approaches represent lower and upper bounds on carbon emission reductions, respectively, and the gap between the two represents the benefits of additional incentives with complete knowledge of human factors. Therefore, the difference in reduction between our learning-augmented approach and the status quo shows the effect of additional incentives with imperfect knowledge of human factors; we term this gap as the \emph{imperfect incentive benefit}. The gap between the optimal incentive allocation and our approach quantifies the cost of imperfect knowledge of human factors; we term this gap as \emph{imperfect knowledge cost}. 

We evaluate the decarbonization performance on a yearly basis using the average percentage reduction in carbon emissions for all households in our data set, computed by dividing the carbon emissions post-installation by the carbon emissions under no intervention, and measure the fidelity in enforcing equity constraints by comparing the incentive allocation across socioeconomic groups. 

\vspace{-0.15cm}
\subsection{Optimizing for carbon reduction}
We first compare the decarbonization performance of various incentive allocation approaches under budgets varying from \$1 million to \$10 million USD and payback periods of 5, 10, and 15 years.

We use the NetBenefit cost model with a discount rate of 5\% in \autoref{fig:costmodel1nonEquity}, and a discount rate of 2\% in \autoref{fig:costmodel2nonEquity}.
 
Both figures show carbon emissions reductions for all incentive allocation approaches. 
The optimal incentive allocation and the status quo set the upper and lower bounds on the reduction of carbon emissions, respectively. Under the status quo in \autoref{fig:costmodel1nonEquity}, the carbon reduction is roughly 3\% when the payback period is 5 years. This is expected as even the houses that benefit the most from decarbonization cannot recoup their investment quickly. As the payback period increases to 15 years, the lower bound on savings increases to 74.91\%.

Our proposed Bandit Incentive Allocation approach lies between the two bounds on carbon reductions. The improvement in reductions reaches a maximum of 31.23\% in a 5-year payback period, 31.36\% in a 10-year payback period and 11.23\% in a 15-year payback period. We observe that the imperfect knowledge cost (the gap between our approach and upper bound) increases as the budget increases; it is 10.06\% at a 1 million USD budget and 32.23\% at a 10 million budget for a 10-year payback period. This shows that the knowledge of human factors significantly affects how well the budget is utilized.

\autoref{fig:costmodel2nonEquity} evaluates various approaches under a discount model corresponding to the high growth rate in energy costs and inflation. We observe similar performance trends to the previous setup. However, carbon reductions are higher in this scenario as more homes adopt decarbonization technologies -- this is because low discount factors means that future returns have a higher value, tipping the cost-benefit ratio for additional households and resulting in a broader acceptance of incentives. 

\begin{figure}[t]
    \centering
    \includegraphics[width=0.8\linewidth]{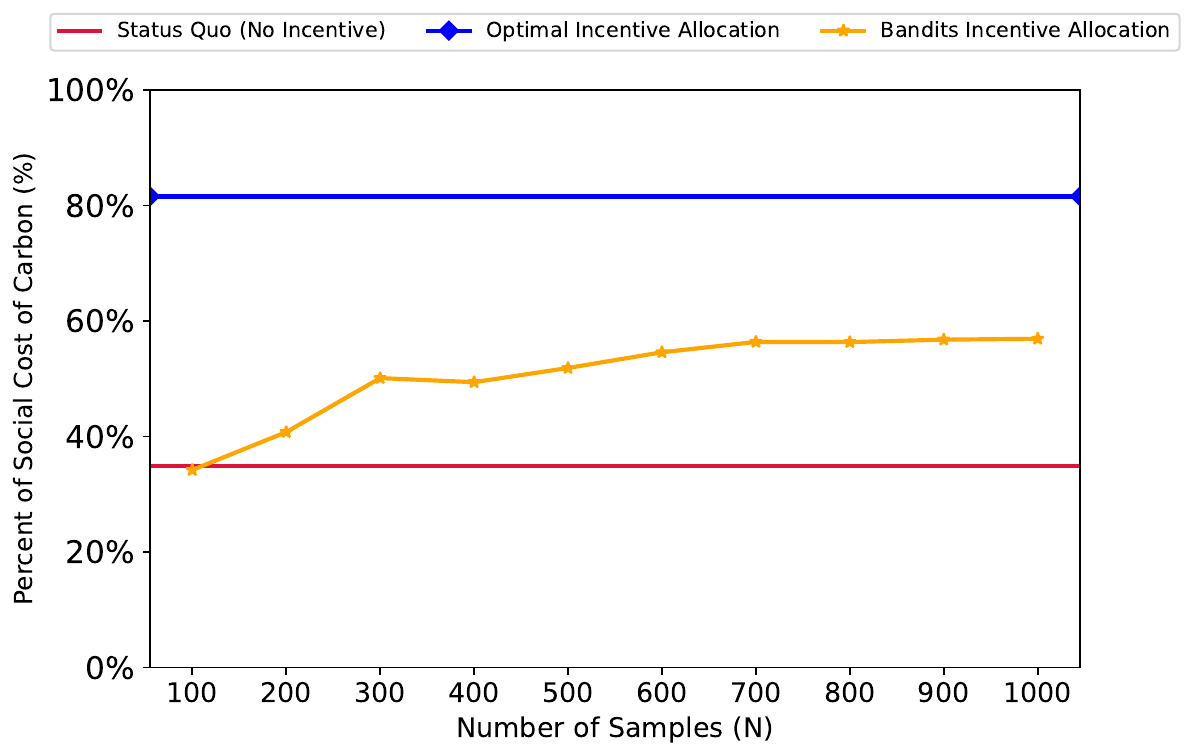}
    \vspace{-0.45cm}
    \caption{Performance of Bandits Incentive Allocation with a budget of \$5 million USD as survey size $N$ varies.}  \label{fig:samples}
    \label{fig:samplefig}
\end{figure}

We note that in a real-world scenario, it might be desirable for the incentive designer to survey as few households as possible during the survey phase -- in \autoref{fig:samplefig}, we evaluate the carbon reduction of the Bandit Incentive Allocation strategy for a single budget of \$5 million USD, varying the survey size $N$.  We observe that in our problem setup, at least 700 homes must be surveyed to reduce the imperfect knowledge cost and approach the best possible carbon reduction.

\vspace{-0.25cm}
\subsection{Incentives under changing environments}

We next analyze the effect of economic and generation parameters such as natural gas prices, electricity prices, and carbon intensity. We set the budget to \$5 million and the payback period to 10 years. 

\autoref{fig:environments}(a) shows the carbon emissions reduction when the future natural gas prices range from \$0.7315/CC to \$2.1/CC. We use the \texttt{ISO-NE} carbon intensity trace and the default electricity price of \$0.14072/kWh.  We note that the ``status quo'' baseline intuitively obtains more carbon reduction as prices go up since finances with existing tax credits and rebates favor the full appliance replacement decarbonization package more often. The Bandit Incentive Allocation achieves an average of 21.23\% additional reductions on the status quo approach and achieves up to 96.17\% as much carbon reduction as the upper bound.
 
\begin{figure}[t]
    \centering
    \includegraphics[width=0.8\linewidth]{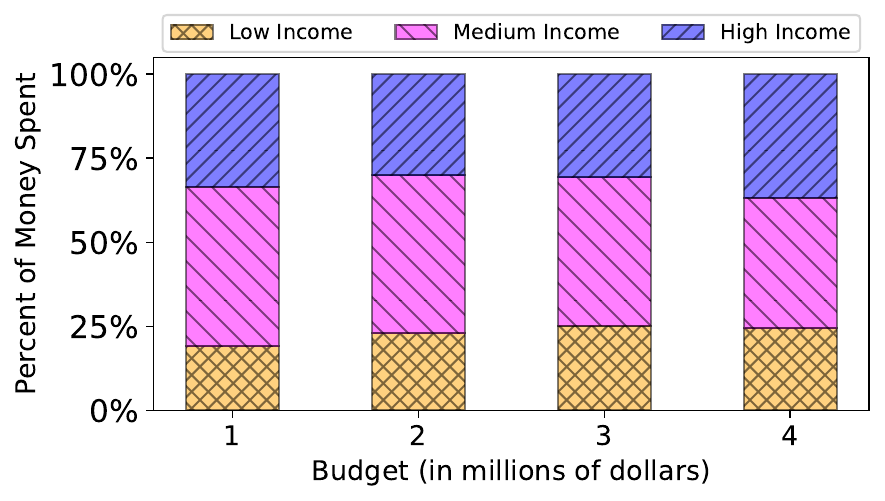}
    \vspace{-0.5cm}
    \caption{The allocation of incentive by the Bandits Incentive Allocation model to high, medium, and low income homes for a few total budget values. } \label{fig:equity}
    \vspace{-0.5cm}
\end{figure}

\begin{figure*}[t]
    \centering
    \includegraphics[width=\linewidth]{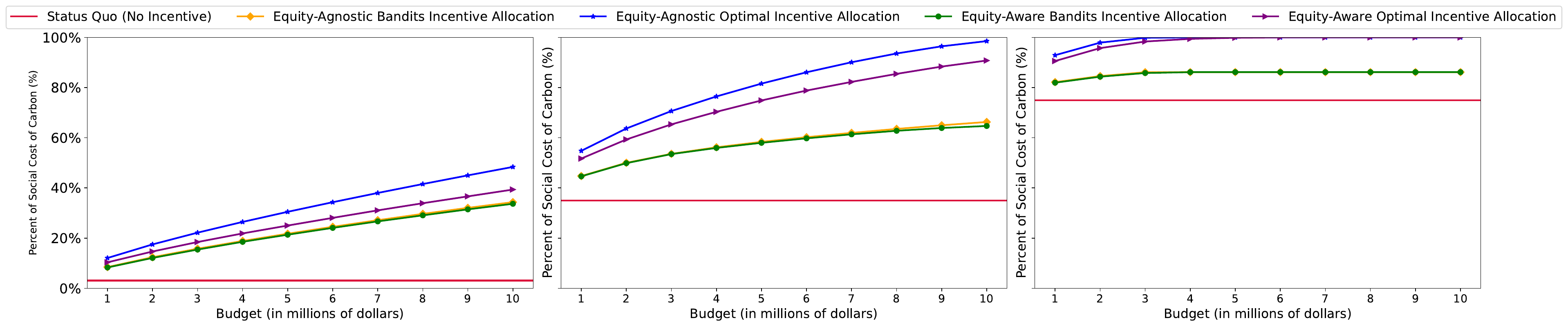}
    \vspace{-0.45cm}
    \caption{Performance of equity-aware variants of optimal and learning-based incentive allocation approaches using a discount rate of 5\%. 
     The equity-aware variant shows a comparable performance to equity-agnostic implementation. } \label{fig:costmodel1equity}
    \vspace{-1em}
\end{figure*}

\autoref{fig:environments}(b) shows the carbon emissions reduction when the future electricity prices range from \$0.077/kWh to \$0.2853/kWh.  We use the \texttt{ISO-NE} carbon intensity trace and the default gas price of \$1.160/CC. 
Unlike gas pricing, the ``status quo'' baseline obtains more carbon reduction as prices go up to a certain extent due to the improved returns with fixed investment. The additional carbon reduction of our learning-driven approach is 22.86\% on average across all prices compared to the status quo.  It also achieves a maximum of 87.33\% of the reductions of the optimal approach.

Finally, \autoref{fig:environments}(c) shows carbon emissions reduction across five different grid carbon intensity traces. The gas and electricity prices are fixed to \$1.160/CC and \$0.14072/kWh, respectively. 
We observe a gradual decline in carbon reduction as the grids become ``cleaner''. This is because decarbonization packages reduce grid reliance -- transitioning away from a dirty grid makes a much larger impact on carbon savings if the existing grid is dirtier. While the upper bound of carbon reduction varies, the Bandit Incentive Allocation is able to achieve significantly higher carbon savings than the status quo, demonstrating the efficacy of our approach across regions irrespective of electric grid characteristics.

\subsection{Optimizing equity and decarbonization} \label{sec:evalEq}
As shown by prior work~\cite{Sunter:19, Brown2022, Crago:23, PobleteCazenave2023}, incentives can exacerbate socioeconomic inequities.  In this experiment, we first examine the equity impacts of our approach. \autoref{fig:equity} shows the split of incentive budget across the low, medium, and high-income groups, showing a clear deviation from the equitable distribution. In the following experiments, we impose an equity constraint such that the fraction of investment in the low, medium, and high income groups corresponds to 25\%, 50\%, and 25\%. However, the relative ratio is configurable and can be set by an exogenous policy adopting a different equity definition.

\autoref{fig:costmodel1equity} compares the carbon reduction performance of equity-aware incentive allocation schemes (Optimal and the Bandit Incentive Allocation), their equity-agnostic counterparts (shown in \autoref{fig:costmodel1nonEquity}) against the equity-agnostic status quo (NetBenefit cost model with a discount rate of 5\%). The equity-aware optimal incentive allocation approach shows a slightly lower carbon reduction across all budgets and payback periods, illustrating a manageable trade-off between carbon emission reductions and equity implications. We find that the Bandit Allocation strategy is robust to the equity-aware constraints, as the Equity-Aware allocation is always within 2\% of the Equity-Agnostic allocation across all payback periods. However, we do observe that there is a smaller difference between these two Bandit Allocation strategies when the budget is small.  This is because as the budget increases, equity restrictions begin to impact the achievable reductions by limiting the number of additional homes that can be selected.

\subsection{Incentives over time}
\begin{figure}[t]
    
    \centering
    \includegraphics[width=1\linewidth, scale=1.5]{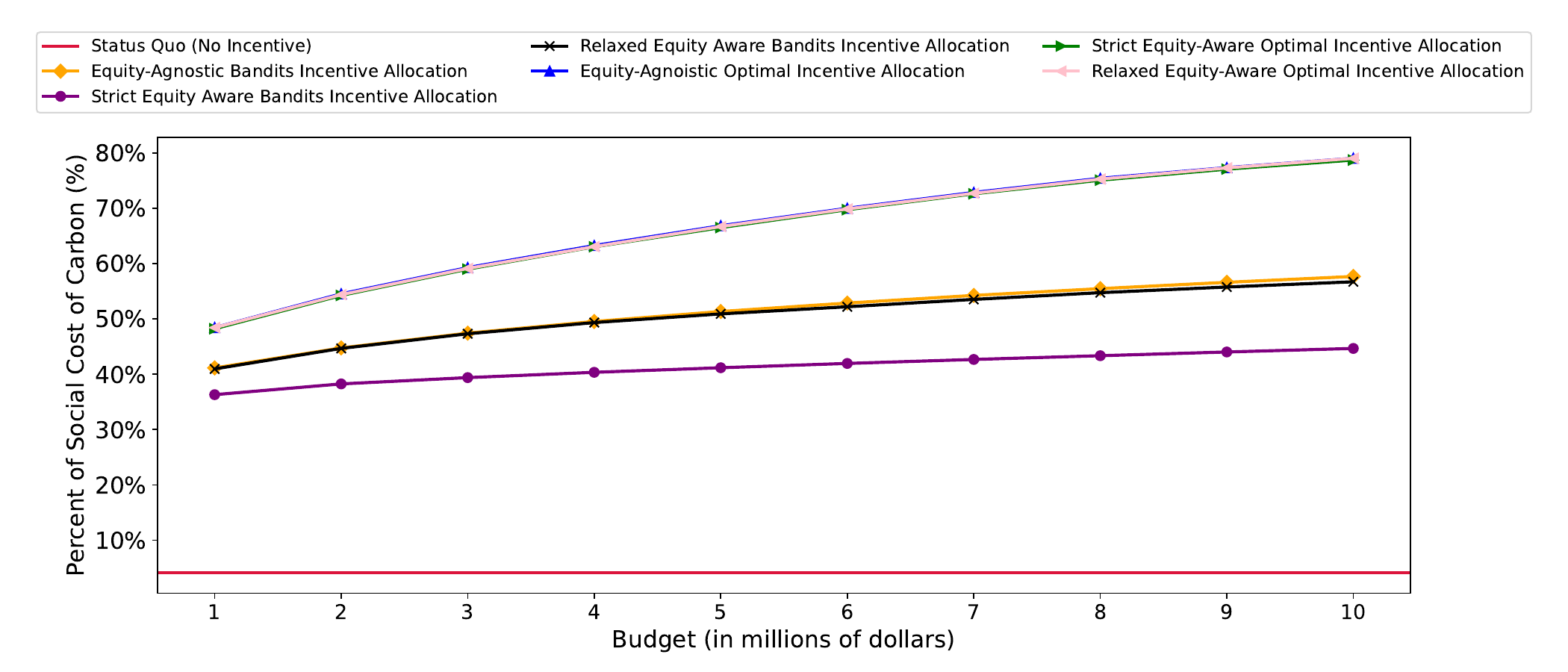}
    \vspace{-0.45cm}
    \caption{Carbon reduced over a span of 10 years including equity-aware strategies}  \label{fig:overtime}
    
    \vspace{-1em}
\end{figure}

\begin{figure}[t]
    \centering
    \includegraphics[width=0.8\linewidth]{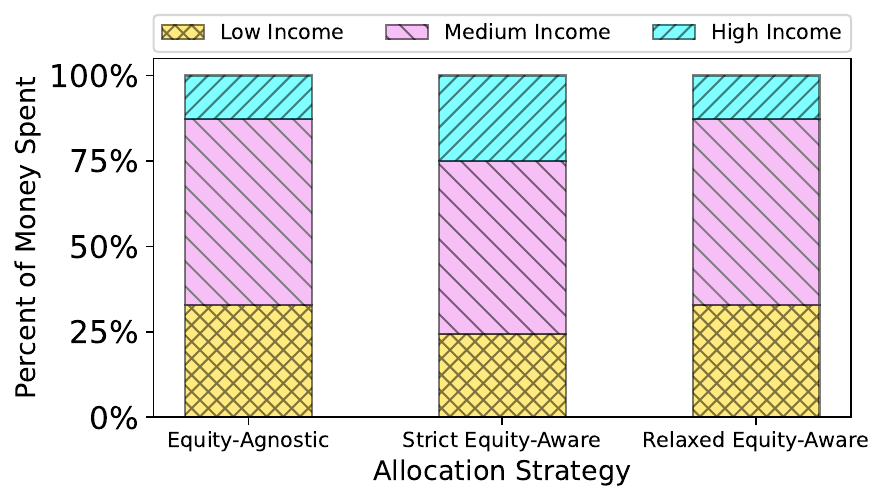}\\
    \vspace{-0.2cm}
    \textbf{a) Allocation of budget = 1 million USD}
    \includegraphics[width=0.8\linewidth]{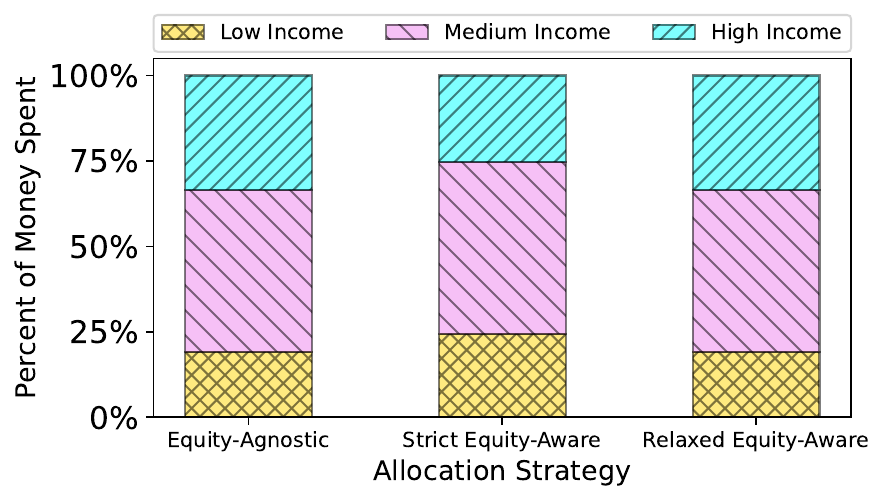}\\
    \textbf{b) Allocation of budget =10 million USD }
    \vspace{-1em}
    \caption{\normalsize Bandit Incentive Allocations to high, medium and low-income homes during the first year of incentives.} \label{fig:overtimepercent}
   
    \vspace{-0.5cm}
\end{figure}

Offering incentives in stages over long periods of time has the potential to further exacerbate inequities, as homes that are decarbonized earlier are able to realize more benefits from decarbonization. In these experiments, we examine how our approach is impacted by the constraint described in \autoref{eq:consot}, where incentives are offered over many years, but only a fraction of the budget can be used yearly. For this scenario, we impose two different equity constraints (Strict-Equity and Relaxed Equity) on the optimization problem, which are described by \autoref{eq:consStrictEquity} and \autoref{eq:consRelaxedEquity}. We divide the total incentive budget equitably among high, middle and low income groups, such that the fraction of the investment in the low, medium and high income groups are 25\%, 50\% and 25\% respectively. 
We use the NetBenefit cost model with a discount rate of 5\% and payback period of 10 years. We allow incentives to be allocated for a total of 10 years. \autoref{fig:overtime} compares the carbon savings of the Relaxed-Equity Aware, Strict-Equity Aware and Equity-Agnostic Bandit Incentive Allocation strategies, as well as the carbon savings for their respective optimal upper bound and lower bound. We observe that the Optimal Incentive Allocation strategies all result in roughly the same carbon savings (within a 1\% difference of each other), with the Strict-Equity allocation resulting in the smallest carbon savings out of all of the optimal schemes. This indicates that with perfect knowledge of the incentives for each home, there is a minimal trade-off between optimizing for equity and optimizing for carbon savings.
Additionally, the Bandits Incentive Allocation strategies all achieve over a 32.16\% increase in carbon savings from the status quo. The Equity-Agnostic strategy achieves up to a 53.52\% gain in carbon savings over the status quo, the Strict-Equity Aware strategy achieves a maximum of 40.51\% over the status quo, and the Relaxed-Equity Aware strategy achieves a maximum gain of 52.56\%. Intuitively, we find that the Relaxed-Equity Aware strategy outperforms the Strict-Equity Aware strategy. 
\autoref{fig:overtimepercent} shows the split of the yearly budget between income levels for Equity-Agnostic, Strict-Equity and Relaxed-Equity strategies after the first year of incentive allocation for a total budget. In \autoref{fig:overtimepercent}(a), it is evident that that the Equity-Agnostic and Relaxed-Equity incentive allocation strategies do not allocate incentives perfectly equitably, with these strategies over-allocating incentives to medium-income and low-income homes. This is because the budget allocated for high-income homes is insufficient to offer acceptable incentives under all scenarios.\footnote{In the usual sense, giving more incentives to medium- and low-income homes would be considered equitable.  However, the purpose of this experiment is to show how well each strategy conforms to a desired distribution that has been deemed ``equitable''.} 
In  \autoref{fig:overtimepercent}(b), where the budget is 10 million USD, we see that  the Equity-Agnostic and Relaxed-Equity incentive allocation strategies now over-allocate to high-income homes. In contrast, the Strict-Equity Aware strategy allocates incentives according to the specified distribution under both budgets. These results introduce a trade-off between carbon and equity, where relaxing equity constraints leads to a less equitable allocation, but more carbon savings. 

\vspace{-0.15cm}
\section{Related Work}
\label{sec:relwork}
In this section, we review prior work that studies the effects of existing incentives, simulates responses to incentive plans, and proposes new incentives toward general residential decarbonization. 

\noindent\textbf{Effects of existing incentives. \ }
There have been numerous studies that investigate the real-world effects of decarbonization incentive programs such as tax credits, rebates, grants, net metering (i.e., feed-in tariffs) and renewable energy credit markets. Perhaps the most well-studied is residential solar adoption~\cite{Sarzynski2012, Bauner2015, Hagerman2016, Crago2017, Matisoff2017, Sunter:19, Boccard2021, Peasco2021, Kearns2022, Crago:23}.  \citet{Matisoff2017} review state and utility incentives for residential PV in the United States, finding that point of sale rebates are up to $8\times$ more effective compared to tax credits worth the same amount.  Other studies come to similar conclusions, including~\cite{Crago2017, Kearns2022}.  \citet{Sunter:19} and \citet{Crago:23} consider the distributional impacts of solar in terms of adoption and financial returns, respectively, finding racial and income disparities.

Other studies have considered the effects of incentives on heat pump and battery storage adoption~\cite{Bjrnstad2012, Snape2015, Alberini2015, Alberini2016, Lill2019, Carroll2020, Brown2022, Shen2022, Davis2023}. A Norway-based case study finds that 54.2\% of participants in a subsidy program were ``very satisfied'' with heat pumps~\cite{Bjrnstad2012}. 

Interestingly, \citet{Davis2023} suggests that heat pump adoption is not well-correlated with income, while \citet{Brown2022} find disparities along racial and income lines in battery storage adoption throughout California. 

Although they do not propose new incentive structures, these studies help contextualize the landscape of incentives for residential decarbonization  and influence our high-level approach.

\noindent\textbf{Incentive simulations and proposals. \ }
Prior work also uses simulations to estimate the adoption of new residential technologies based on existing or proposed incentives.  Many find that upfront subsidies encourage more adoption, including~\cite{Lobel2011, Zhao2012, Hsu2012, Burr2016, Zander2019, Kokoni2021, Sher2022, DAdamo2022, PobleteCazenave2023}.  Others simulate the financial viability and effects of climates on heat pumps or battery storage~\cite{Blumsack2012, Peerapong2014, Agnew2017, Nousdilis2020,Chatterji2020,Vonsien2020, Dougherty2021, Ruffino2022, DAdamo2022}.  Some works raise questions about the payback period of BESS systems in low-solar areas without additional incentives, including~\cite{Jones2017, Zhang2018, Zakeri2021}.  Others simulate the potential of heat pumps, finding that both the economics and the decarbonization potential depend heavily on e.g., electricity generation mix, pricing, and taxation~\cite{Barnes2020, Gaur2021, Kokoni2021,PobleteCazenave2023}.

Other work uses a combination of simulation data, economic analysis, and empirical results to propose new incentive structures for residential decarbonization technologies, including~\cite{Srinivasan2009, Mulder2013, Hannon2015, Patteeuw2016, Sharma2019, Bunea2020, Varghese2020, Guo2021, Tibebu2021, Chattopadhyay2022, Lin2022}.  Most closely related to our work, \citet{Marinakis2018} develops an optimization-based model to design incentives for solar PV, heat pump, and BESS combination systems, although they do not explicitly consider carbon reduction, instead focusing on the adoption rate.  \citet{Vimpari2021} suggest that energy efficiency subsidies should be allocated into areas with lower housing prices, since low-income areas pay relatively more for energy. 
In contrast to  the above studies,  our focus is  on carbon reduction as opposed to adoption, which adds a dimension to our analysis.

\noindent\textbf{Mechanism design and learning. \ }
Our study draws on foundational work in mechanism design and multi-armed bandits.  In the field of mechanism design, our work is most related to dynamic mechanism and incentive design, where the underlying agents (e.g. households) update their action according to some unknown but learnable rules~\cite{Parkes2010, Cai2013, Pothineni2014, Pavan2014, Ratliff2019, Bergemann2019, Ratliff2021}.  
In the broad literature on multi-armed bandits, we mostly draw on the problem of offline contextual bandits, where learners estimate the quality of actions by leveraging a pre-collected data set with contextual information and make informed decisions in uncertain environments~\cite{auer2002nonstochastic,li2010contextual,slivkins2011contextual,agarwal2014taming,foster2018practical,jin2021pessimism,li2022pessimism,rashidinejad2022bridging}. 
This problem has been extensively studied and finds applications in diverse domains, including healthcare systems~\cite{durand2018contextual,bastani2020online}, recommender systems~\cite{li2010contextual,bouneffouf2012contextual}, and cyber-physical systems~\cite{li2016contextual,wu2019context}.
This work applies offline contextual bandits towards the practical problem of allocating incentives for decarbonization.

\vspace{-0.15cm}
\section{Conclusion}
\label{sec:conclusion}
We present a novel data-driven approach for holistic dynamic incentive allocation for city-scale deep decarbonization.  
We defined an optimization model that dynamically allocates a total incentive budget to households to directly maximize the resultant carbon emissions reduction and leveraged learning-based techniques to estimate human factors, such as a household's willingness to adopt new technologies given a certain incentive.
We evaluated our approach using a real data set and showed that our learning-based technique significantly outperforms an example of status quo incentives offered by a state and utility, achieving up to 32.23\% additional carbon reductions, and an average of 78.84\% of the optimal solution carbon reduction even under equity constraints. Further, our incentive allocation approach achieves significant carbon reduction in a variety of environments, with varying values for the grid carbon intensity, gas prices and electric prices. In future work, it would be very interesting to consider how our approach applies to  other residential sector decarbonization problems.



\vspace{-0.15cm}
\bibliographystyle{ACM-Reference-Format}
\bibliography{main}











\end{document}